\documentclass[prl]{revtex4}
\usepackage{graphicx}
\usepackage{dcolumn}
\usepackage{bm}

\begin{document}
\renewcommand{\theequation}{\arabic{equation}}

\title{Classical and Quantum Aspects of Particle Propagation in External Gravitational Fields\\}

\author{Giorgio Papini}
\altaffiliation[Electronic address:]{papini@uregina.ca}
\affiliation{Department of Physics and Prairie Particle Physics
Institute, University of Regina, Regina, Sask, S4S 0A2, Canada}


\begin{abstract}
In the study of covariant wave equations, linear gravity manifests itself through the metric deviation $\gamma_{\mu\nu}$
and a two-point vector potential $K_{\lambda}$ itself constructed from $\gamma_{\mu\nu}$ and its derivatives.
The simultaneous presence of the two gravitational potentials is non contradictory. Particles also assume the character
of quasiparticles and $K_{\lambda}$ carries information about the matter with which it interacts.
We consider the influence of $K_{\lambda}$ on the dispersion relations of the
particles involved, the particles' motion, quantum tunneling through a horizon, radiation, energy-momentum dissipation
and flux quantization. 
\end{abstract}

 \maketitle

\setcounter{equation}{0}

\section{Introduction}               
Fields for which quantum fluctuations are negligible relative to the expectation
values of the fields themselves can be described by classical equations. This is the
external field approximation. In electrodynamics radiation processes are greatly enhanced
when incident photons are replaced by external fields. Processes like photon-electron
collisions with pair production by a Coulomb field and bremsstrahlung owe their relevance
in physics and astrophysics to the external field approximation.

In gravitation where field strengths are much lower than in electromagnetism and means to increase
cross sections are very desirable, external field problems have been paid limited
attention. This is presumably due to the large number of components of the
gravitational potential and to the difficulties of finding manageable modifications
for the basic fields of interest as required by the characteristics of the gravitational
sources, like rotation and time-dependence. A variety of problems involving gravitational fields of weak to
intermediate strength do not require the full-fledged use of general relativity and can
be tackled using an external field approximation. Over the years
solutions of covariant wave equations have been found \cite{PAP0,PAP1,PAP2,PAP3,PAP4} that
are exact to first order in
the metric deviation $\gamma_{\mu\nu}= g_{\mu\nu}-\eta_{\mu\nu}$, where $\eta_{\mu\nu}$ is the Minkowski metric,
and do not depend on the choice of any field equations for $\gamma_{\mu\nu}$. They are a useful tool in the study of
the interaction of gravity with quantum systems
and have been applied to
interferometry and gyroscopy \cite{PAP1}, the optics of particles \cite{PAP3}, the observation
of gravitational effects \cite{PAP1}, the spin-flip of particles under the action of inertial
and gravitational fields \cite{PAP2}, spin currents \cite{PASP}, neutrino physics \cite{PAP5,LAMB} and radiative processes in astrophysics \cite{PAP6,PAP7,PAP8,PAP9}.

In all the solutions mentioned, gravity is contained in a phase factor
whose main ingredient is the two-point \cite{Ruse,Synge} potential $K_{\lambda}(z,x) $
that is a direct consequence of genuinely quantum equations and the expression,
at the same time, of
hidden symmetries of gravity. It is a novel feature of the approach. $K_{\lambda}$ describes gravity
while remaining tied to a metric tensor. That is, $K_{\lambda}$ is known only if $\gamma_{\mu\nu}$ and its derivatives are known.
$K_{\lambda}$ has other interesting aspects that are
studied below. We show, in particular, that
the potential $K_{\lambda}$ affects a particle's dispersion relations, its motion, the WKB problem, energy-momentum dissipation,
radiation and flux quantization.

An interesting development is that as gravity approaches the quantum domain, particles
assume the character of quasiparticles and the gravitational potential carries information
about the matter with which it interacts.

Spin terms that appear in the solutions of the covariant wave equations
are not particularly relevant in what follows, thus we can simply consider the covariant Klein-Gordon (KG) equation.
Similar conclusions can, of course, be reached starting from other known wave
equations, whose solution is based, in any case, on the KG equation.

Neglecting curvature dependent terms and applying the Lanczos-De Donder condition
\begin{equation}\label{L}
  \gamma_{\alpha\nu,}^{\,\,\,\,\,\,\,
  \nu}-\frac{1}{2}\gamma_{\sigma,\alpha}^\sigma = 0 \,,
  \end{equation}
we can write the covariant KG equation to $\mathcal{O}(\gamma_{\mu\nu})$ in the form
\begin{equation}\label{KG}
\left(\nabla_{\mu}\nabla^{\mu}+m^2\right)\phi(x)\simeq\left[\eta_{\mu\nu}\partial^{\mu}\partial^{\nu}+m^2
+\gamma_{\mu\nu}\partial^{\mu}\partial^{\nu}
\right]\phi(x)=0\,.
\end{equation}
Units $\hbar=c=k_{B}=1$ are used unless specified otherwise. The notations are as in \cite{PAP7}.
In particular, partial derivatives with respect to a variable $x_{\mu}$ are interchangeably
indicated by $\partial_{\mu}$, or by a comma followed by $\mu$.

The first order solution of (\ref{KG}) is
\begin{equation}\label{PHA}
\phi(x)=\left(1-i\Phi_{G}(x)\right)\phi_{0}(x)\,,
\end{equation}
where $\phi_{0}(x)$ is a plane wave solution of the free KG equation
\begin{equation}\label{KG0}
\left(\partial_{\mu}\partial^{\mu}+m^2\right)\phi_{0}(x)=0 \,,
\end{equation}
and
\begin{equation}\label{PHI}
\Phi_{G}(x)=-\frac{1}{2}\int_P^x
dz^{\lambda}\left(\gamma_{\alpha\lambda,\beta}(z)-\gamma_{\beta\lambda,\alpha}(z)\right)
\left(x^{\alpha}-z^{\alpha}\right)k^{\beta}
\end{equation}
\[+\frac{1}{2}\int_P^x dz^{\lambda}\gamma_{\alpha\lambda}(z)k^{\alpha}=\int_{P}^{x}dz^{\lambda}K_{\lambda}(z,x)\,,\]
where $P$ is an arbitrary point, henceforth dropped, and
\begin{equation}\label{K}
K_{\lambda}(z,x)=-\frac{1}{2}\left[\left(\gamma_{\alpha\lambda,\beta}(z)
-\gamma_{\beta\lambda,\alpha}(z)\right)\left(x^{\alpha}-z^{\alpha}\right)-\gamma_{\beta\lambda}(z)\right]k^{\beta}\,.
\end{equation}
The momentum of the plane wave solution $\phi_0$ of (\ref{KG0})
is $k^{\alpha}$ and satisfies the equation $k_{\alpha}k^{\alpha}=m^2$.
There are no additional constraints on the solution of (\ref{KG}) represented by (\ref{PHA})-(\ref{K}) except
for the order of approximation in $\gamma_{\mu\nu}$. Higher order solutions can be calculated following the procedure outlined in
\cite{PAP7}. Nonetheless our calculations are limited to $O(\gamma_{\mu\nu})$ as we are primarily concerned with gravitational fields of weak to
moderate intensity.

It is easy to see that (\ref{PHA}) is a solution of (\ref{KG}). By writing $\phi\equiv\phi^{(1)}$ for the
first order solution and differentiating (\ref{PHA}) with respect to $x_{\mu}$, we obtain
\begin{equation}\label{diff1}
\phi_{,\mu}^{(1)}=\phi_{0,\mu}-i\Phi_{G,\mu}\phi_{0}-i\Phi_{G}\phi_{0,\mu}\,,
\end{equation}
and again
\begin{equation}\label{diff2}
\phi_{,\mu\nu}^{(1)}=\phi_{0,\mu\nu}-i\Phi_{g,\mu\nu}\phi_{0}-i\Phi_{G,\mu}\phi_{0,\nu}-i\Phi_{G,\nu}\phi_{0,\mu}-i\Phi_{G}\phi_{0,\mu\nu}\,.
\end{equation}
The result then follows by substituting (\ref{diff2}) in (\ref{KG}) and by noticing that $\eta^{\mu\nu}\Phi_{G,\mu\nu}=\eta^{\mu\nu}k_{\alpha}\Gamma^{\alpha}_{\mu\nu}=0$
by virtue of (\ref{L}) and $k^{\mu}\Phi_{G,\mu}=\frac{1}{2}\gamma^{\mu\nu}k_{\mu}k_{\nu}$.

\section{Quasiparticles}

Equations (\ref{PHA}), (\ref{PHI}) and (\ref{K})  are the byproduct of
covariance (minimal coupling) and, ultimately, of Lorentz
invariance and can therefore be applied to general relativity, in
particular to theories in which acceleration has an upper limit
\cite{CAI1,CAI2,CAI3,BRA,MASH1,MASH2,MASH3,TOLL} and that therefore allow the resolution of
astrophysical \cite{SCHW,RN,KERR,PU} and cosmological singularities
in quantum theories of gravity \cite{ROV,BRU}. They also
are  relevant to those theories of asymptotically safe gravity
that can be expressed as Einstein gravity coupled to a scalar
field \cite{CA}.

Path-dependent field variables in electromagnetism and gravitation have been used in the works of
Volkov \cite{VOL}, Bergmann \cite{BER}, DeWitt \cite{DEW} and Mandelstam \cite{MAND}.
From (\ref{PHA}) we can derive expressions for the covariant derivatives of path-dependent quantities.
The vector field $K_{\lambda}$ plays a role similar to that of the vector
potential in electromagnetism, but contains, at the same time, reference to matter through the momentum $k_\mu$ of $\phi_0$.
It is, in fact, this association that suggests
the introduction of the notion of quasiparticle that in other areas of physics
describes fields and particles whose properties are affected by the presence of
other particles and media with which they interact.

$K_\lambda(z,x)$ also satisfies Maxwell-type equations identically.
By differentiating (\ref{K}) with
respect to $z^{\alpha}$, we find \cite{PAP10}
\begin{equation}\label{FT}
\tilde{F}_{\mu\lambda}(z,x)=K_{\lambda,\mu}(z,x)-K_{\mu,\lambda}(z,x)=R_{\mu\lambda\alpha\beta}(z) J^{\alpha\beta}\,,
\end{equation}
where
$R_{\alpha\beta\lambda\mu}(z)=-\frac{1}{2}\left(\gamma_{\alpha\lambda,\beta\mu}
+\gamma_{\beta\mu,\alpha\lambda}-\gamma_{\alpha\mu,\beta\lambda}-\gamma_{\beta\lambda,\alpha\mu}\right)$
is the linearized Riemann tensor satisfying the identity
$R_{\mu\nu\sigma\tau}+R_{\nu\sigma\mu\tau}+R_{\sigma\mu\nu\tau}=0$
and
$J^{\alpha\beta}=\frac{1}{2}\left[\left(x^{\alpha}-z^{\alpha}\right)k^\beta-k^\alpha
\left(x^\beta-z^\beta\right)\right]$ is the angular momentum about
the base point $x^\alpha$.
Maxwell-type equations
\begin{equation}\label{ME1}
\tilde{F}_{\mu\lambda,\sigma}+\tilde{F}_{\lambda\sigma,\mu}+\tilde{F}_{\sigma\mu,\lambda}=0
\end{equation}
and
\begin{equation}\label{ME2}
\tilde{F}^{\mu\lambda}_{\,\,\,\,\,\,\,,\lambda}\equiv j^{\mu}=
\left(R^{\mu\lambda}_{\,\,\,\,\,\,\,\alpha\beta}J^{\alpha\beta}\right),_{\lambda}
=R^{\mu\lambda}_{\,\,\,\,\,\,\,\alpha\beta,\lambda}\left(x^\alpha
-z^\alpha\right)k^\beta +R^{\mu}_{\,\,\,\,\beta}k^{\beta}\,,
\end{equation}
can be obtained from (\ref{FT}) using the Bianchi identities $R_{\mu\nu\sigma\tau,\rho}
+R_{\mu\nu\tau\rho,\sigma}+R_{\mu\nu\rho\sigma,\tau}=0$. The current $j^{\mu}$ satisfies the conservation law
$j^{\mu}_{\,\,\,,\mu}=0$.
Equations (\ref{ME1}) and (\ref{ME2}) are identities and do not represent additional constraints on $\gamma_{\mu\nu}$.
One finds, in particular, that the "electric" and "magnetic" components of $\tilde{F}_{\mu\nu}$ are
\begin{equation}\label{EM}
\tilde{E_{i}}=R_{0i\alpha\beta}J^{\alpha\beta}\,\,\,,\tilde{H_{i}}=\epsilon_{ijk}R^{kj}_{\,\,\,\,\,\,\alpha\beta}J^{\alpha\beta}\,,
\end{equation}
where $\epsilon_{ijk}$ is the Levi-Civita symbol.

In this work we investigate some of the consequences that follow from (\ref{ME1}) and (\ref{ME2}) and
from the close association of $K_{\lambda}$ with matter.

The recombination of ten $\gamma_{\mu\nu}$ into four $K_{\lambda}$ is a remarkable
phenomenon, even though knowledge of all $\gamma_{\mu\nu}$ is still needed, in general, to calculate $K_\lambda$ \cite{PAP10}.
It follows from (\ref{K}) that
the gravitational field is described by $K_{\lambda}$ along a particle world line and that
$K_\lambda$ vanishes wherever $k_{\mu}$ does.
From (\ref{ME2}) we also find that
$j^{\mu}=-\frac{1}{2}(\partial^2 \gamma^{\mu}_{\alpha,\beta}-\partial^2 \gamma^{\mu}_{\beta,\alpha})(x^{\alpha}-z^{\alpha})k^{\beta}+
\frac{1}{2}\partial^2 \gamma^{\mu}_{\beta}k^{\beta}$. The current $j_{\mu}$ vanishes when $\partial^2 \gamma_{\mu\beta}=0$ that corresponds
to pure gravitational fields because, in the linear approximation, $R_{\mu\beta} =\frac{1}{2}\partial^2 \gamma_{\mu\beta}=0$.
When $j_{\mu}=0$, equations (\ref{ME1}) and (\ref{ME2}) become scale invariant and so does the field $K_{\lambda}$.
In fact, by differentiating
(\ref{K}), we obtain
\begin{equation}\label{DK}
\partial^2 K_{\lambda}=-\frac{k^{\beta}}{2}\left[\left(\partial^2 (\gamma_{\alpha\lambda,\beta})-\partial^2(\gamma_{\beta\lambda,\alpha})\right)\left(x^\alpha -z^\alpha\right)+
\partial^2 \gamma_{\beta\lambda}-\gamma_{,\lambda\beta}\right]\,.
\end{equation}
The last term in (\ref{DK}) can be eliminated by a gauge transformation.
We then obtain $\partial^{2}K_{\lambda}=0$, irrespective of the value of $k_{\alpha}$.

\section{Dispersion relations and particle motion}

By using Schroedinger's logarithmic transformation \cite{LANC} $\phi = e^{-iS} $ we can pass from the KG equation (\ref{KG}) to the quantum Hamilton-Jacobi
equation. We find to first order in $\gamma_{\mu\nu}$
\begin{equation}\label{QHJ}
i(\eta^{\mu\nu}-\gamma^{\mu\nu})\partial_{\mu}\partial_{\nu}S-(\eta^{\mu\nu}-\gamma^{\mu\nu})\partial_{\mu}S\partial_{\nu}S +m^2=0 \,,
\end{equation}
where
\begin{equation}\label{S}
S= k^{\beta}\left\{x_{\beta}+\frac{1}{2}\int^{x}dz^{\lambda}\gamma_{\beta\lambda}(z)-\frac{1}{2}\int^{x}dz^{\lambda}\left(\gamma_{\alpha\lambda,\beta}(z)-\gamma_{\beta\lambda,\alpha}(z)\right)
\left(x^{\alpha}-z^{\alpha}\right)\right\}
\end{equation}
\[\equiv k_{\alpha}x^{\alpha}+A+B \,.\]
It is well-known that the Hamilton-Jacobi equation is equivalent to Fresnel's wave equation in the limit of large
frequencies \cite{LANC}. However, at smaller, or moderate frequencies the complete  equation (\ref{QHJ}) should be used. We follow
this path. By substituting (\ref{PHA}) into the first term of (\ref{QHJ}), we obtain
\begin{equation}\label{ldd}
i(\eta^{\mu\nu}-\gamma^{\mu\nu})\partial_{\mu}\partial_{\nu}S=i\eta^{\mu\nu}\partial_{\mu}(k_{\nu}+\Phi_{G,\nu})-i\gamma^{\mu\nu}\partial_{\mu}k_{\nu}=i\eta^{\mu\nu}\Phi_{G,\mu\nu}=i
k_{\alpha}\eta^{\mu\nu}\Gamma^{\alpha}_{\mu\nu}=0 \,,
\end{equation}
on account of (\ref{L}). This part of (\ref{QHJ}) is usually neglected in the limit $\hbar\rightarrow 0$.
Here it vanishes as a consequence of solution (\ref{PHA}). The remaining terms of (\ref{QHJ}) yield the classical Hamilton-Jacobi equation
\begin{equation}\label{CHJ}
(\eta^{\mu\nu}-\gamma^{\mu\nu})\partial_{\mu}S\partial_{\nu}S -m^2=\gamma^{\mu\nu}k_{\mu}k_{\nu}-2k^{\mu}\Phi_{G,\mu}=0 \,,
\end{equation}
because $k^{\mu}\Phi_{G,\mu}=1/2 \gamma^{\mu\nu}k_{\mu}k_{\nu}$. Equation (\ref{PHA}) is  therefore a solution of the more general quantum
equation (\ref{QHJ}). It also follows that the particle acquires a generalized "momentum"
\begin{equation}\label{mom}
P_{\mu}=k_{\mu}+\Phi_{G,\mu}=k_{\mu}+\frac{1}{2}\gamma_{\alpha\mu}k^{\alpha}-\frac{1}{2}\int^{x}dz^{\lambda}\left(\gamma_{\mu\lambda,\beta}(z)-
\gamma_{\beta\lambda,\mu}(z)\right)k^{\beta}\,,
\end{equation}
that satisfies the dispersion relation
\begin{equation}\label{eff}
P_{\mu}P^{\mu}\equiv m_{e}^2 = m^2\left(1+\gamma_{\alpha\mu}(x)u^{\alpha}u^{\mu}-\frac{1}{2}\int^{x}dz^{\lambda}\left(\gamma_{\mu\lambda,\beta}(z)-
\gamma_{\beta\lambda,\mu}(z)\right)u^{\mu}u^{\beta}\right)\,.
\end{equation}
The integral in (\ref{eff}) vanishes because $u^{\mu}u^{\beta}$ is contracted on the antisymmetric tensor in round brackets. The effective
mass $m_{e}$ is not in general constant.
In this connection too we can speak of quasiparticles. The medium in which the scalar particles propagate is here represented by space-time.

We have already used some of the properties of
(\ref{mom}) and (\ref{eff}) elsewhere \cite{PAP4,PAP8}.
$P_{\mu}$ of (\ref{mom}) describes the geometrical optics of particles correctly and gives the correct deflection predicted by general relativity.
On using the relations
\begin{equation}\label{tool1}
\Phi_{G,\mu}=K_{\mu}(x,x)+\int^{x}dz^{\lambda}\partial_{\mu}K_{\lambda}(z,x)\,,
\end{equation}
and
\begin{equation}\label{tool2}
\Phi_{G,\mu\nu}=K_{\mu,\nu}(x,x)+\partial_{\nu}\int^{x}dz^{\lambda}\partial_{\mu}K_{\lambda}(z,x)=k_{\alpha}\Gamma_{\mu\nu}^{\alpha}\,,
\end{equation}
and by differentiating (\ref{mom}) we obtain the covariant derivative of $P_{\mu}$
\begin{equation}\label{acc}
\frac{DP_{\mu}}{Ds}=m\left[\frac{du_{\mu}}{ds}+\frac{1}{2}\left(\gamma_{\alpha\mu,\nu}-\gamma_{\mu\nu,\alpha}+\gamma_{\alpha\nu,\mu}\right)u^{\alpha}u^{\nu}\right]
\end{equation}
\[=m\left(\frac{du_{\mu}}{ds}+\Gamma_{\alpha,\mu\nu}u^{\alpha}u^{\nu}\right)=
\frac{Dk^{\mu}}{Ds}\,.\]
This result is independent of any choice of field equations for $\gamma_{\mu\nu}$. We see from (\ref{acc}) that, if $k_{\mu}$ follows a geodesic, then $\frac{DP_{\mu}}{Ds}=0$  and  $\frac{Dm_{e}^2}{Ds}=0$.
The classical equations of motion are therefore contained in (\ref{acc}), but it would require the particle described by (\ref{KG}) to just choose a geodesic, among all the
paths allowed to a quantum particle.

We also obtain, from (\ref{QHJ}), $\sqrt{(\partial_{i}S)^2}=\pm \sqrt{-m^2+(\partial_{0}S)^2 -\gamma^{\mu\nu}\partial_{\mu}S\partial_{\nu}S}$ which, in the absence of gravity,
gives $k_{i}^2 =\sqrt{-m^2+k_{0}^2}$, as expected. Remarkably, (\ref{mom}) is an exact integral of (\ref{acc})
which can itself be integrated exactly to give the particle's motion
\begin{equation}\label{U}
X_{\mu}=x_{\mu}+\frac{1}{2}\int^{x}dz^{\lambda}\gamma_{\mu\lambda}\left\{\left(\gamma_{\alpha\lambda,\mu}-\gamma_{\mu\lambda,\alpha}\right)\left(x^{\alpha}-z^{\alpha}\right)\right\}\,.
\end{equation}

Higher order approximations to the solution of (\ref{KG}) can be obtained by writing
\begin{equation}\label{IT}
\phi(x)=\Sigma_{n}\phi_{(n)}(x)= \Sigma_{n}e^{-i\hat{\Phi}_{G}} \phi_{(n-1)}\,,
\end{equation}
where the operator $\hat{\Phi}_{G}$ is
\begin{equation}\label{PHIh}
\hat{\Phi}_{G}(x)=-\frac{1}{2}\int_P^x
dz^{\lambda}\left(\gamma_{\alpha\lambda,\beta}(z)-\gamma_{\beta\lambda,\alpha}(z)\right)
\left(x^{\alpha}-z^{\alpha}\right)\hat{k}^{\beta}
\end{equation}
\[+\frac{1}{2}\int_P^x dz^{\lambda}\gamma_{\alpha\lambda}\hat{k}^{\alpha}\,,\]
and $\hat{k}^{\alpha}=i\partial^{\alpha}$.

The solution (\ref{IT}) plays a dynamical role akin to Feynman's path integral formula \cite{FEY}. In (\ref{IT}), however,
it is the solution itself that is varied by successive approximations, rather than the particle's path.

\section{The gravitational WKB problem}

We now study the propagation
of a scalar field in a gravitational background. We know, from standard quantum mechanics \cite{BALL}, that $S$
develops an imaginary part when the particle tunnels through a potential. This imaginary contributions is interpreted as
the transition amplitude across the classically forbidden region, which is therefore given by \cite{AKM}
\begin{equation}\label{TAG}
\mathcal{T}=\exp\left[-2Im(S)\right]=exp\left\{-2Im\left[\ln\left(\Sigma_{n}\exp\left(-i\hat{\Phi}_{G}\phi_{n-1}\right)\right)\right]\right\}\,.
\end{equation}
To $O(\gamma_{\mu\nu})$, (\ref{TAG}) becomes
\begin{equation}\label{TA}
\mathcal{T}=
exp\left\{-2Im\left[ x_{\beta}+
\frac{1}{2}\oint dz^{\lambda}\gamma_{\beta\lambda}(z)
-\frac{1}{2}\oint dz^{\lambda}(\gamma_{\alpha\lambda,\beta}(z)-\gamma_{\beta\lambda,\alpha}(z))(x^{\alpha}-z^{\alpha})\right]k^{\beta}\right\} \,,
\end{equation}
for a space-time path traversing the gravitational background from $-\infty$ to $+\infty$ and back as it must in order to make
(\ref{TA}) invariant. Assuming a Boltzmann distribution
for the particles $\mathcal{T}=e^{-k_{0}/T}$, where $T$ is the temperature and the Boltzmann constant $k_{B}=1$, we find, in general coordinates,
\begin{equation}\label{TU}
T=k_{0} /Im\left\{2k^{\beta}\left[ x_{\beta}+
\frac{1}{2}\oint dz^{\lambda}\gamma_{\beta\lambda}(z)-\frac{1}{2}\oint dz^{\lambda}(\gamma_{\alpha\lambda,\beta}(z)-\gamma_{\beta\lambda,\alpha})(x^{\alpha}-z^{\alpha})\right]\right\}\,.
\end{equation}
The intended application here is to the propagation problem in Rindler
space given by
\begin{equation}\label{R}
ds^{2} =\left(1+a x\right)^2 (dx^{0})^{2} -(dx)^{2}\,,
\end{equation}
with a horizon at $x=-1/a$, where $a^{2}=a_{\alpha}a^{\alpha}$ is the constant proper acceleration measured
in the rest frame of the Rindler observer.
We note that, a priori, our approach is ill-suited to treat this problem that frequently in the literature
is tackled starting from exact, or highly symmetric solutions of the KG equation \cite{TAK}.
In fact the external field approximation $|\gamma_{\mu\nu}|<|\eta_{\mu\nu}|$ may become inadequate close to the horizon, from where
the imaginary part of $\mathcal{T}$ comes, for some systems of coordinates. This requires attention, as discussed below.
Nonetheless the external field approximation has
interesting features like the presence of $k_{\alpha}$ in (\ref{TU}) and manifest covariance and invariance under canonical
transformations.

It is convenient, for our purposes, to use the Schwarzschild-like form for (\ref{R})
using the transformation \cite{DEG}
\begin{equation}\label{TR1}
x^{0}=\frac{1}{a}\sqrt{1+2ax'}\sinh(at')\,,\,\, x^{1}=\frac{1}{a}\sqrt{1+2ax'}\cosh(at')
\end{equation}
for $x^{1}\geq -1/2a$ and the same transformation with the hyperbolic functions interchanged
for $x^{1} \leq -1/2a $. The resulting metric is
\begin{equation}\label{RS}
ds^{2} = \left(1+2ax'\right) (dt')^{2}-\frac{1}{1+2ax'}(dx')^{2}\,,
\end{equation}
for which the horizon is at $x'=-1/2a$. From (\ref{RS}) we find $\gamma_{00}=2ax', \gamma_{11}=2ax'/(1+2ax')$. If
$\gamma_{00}$ and $\gamma_{11}$ represent corrections to the Minkowski metric, we must have $|\gamma_{00}/g_{00}|<1,|\gamma_{11}/g_{11}|<1$
for any $a>0$. The external field approximation therefore remains valid for $-1/4a <x'<1/2a$.
This is sufficient to justify our calculation.
We now write the terms $A$ and $B$, defined in (\ref{S}), for the metric (\ref{RS}) explicitly. We find
\begin{equation}\label{A}
A= \frac{k^{0}}{2}\int dz^{0}\gamma_{00}+\frac{k^{1}}{2}\int dz^{1}\gamma_{11}\,,
\end{equation}
and, by taking the reference point $x^{\mu}=0$,
\begin{equation}\label{B}
B=-\frac{k^{0}}{2}\int dz^{0}\gamma_{00,1}z^{1}+\frac{k^{1}}{2}\int dz^{0}\gamma_{00,1}z^{0}\,.
\end{equation}
The explicit expressions for $A$ and $B$ confirm the fact that $\mathcal{T}$ receives contributions from
both time and space parts of $S$ as pointed out in \cite{DEG}. This is, on the other hand, expected
of a fully  covariant approach.

The first integrals in $A$ and $B$ cancel each other.
The second integral in $A$ can be calculated by contour
integration by writing $z^{1}=-1/2a +\epsilon e^{i\theta}$. The result
$Im\int_{-\infty}^{\infty} dz^{1}z^{1}k^{1}/(1+2az^{1})=-k^{1}\pi/4a^2$ yields a vanishing contribution because $k^{1}$ reverses its sign
on the return trip. The last integral in $B$ is real.
The term $k_{0}\Delta t'$ in (\ref{S}) contributes
the amount $k^{0}(-i\pi/2a)2$ because for a round trip the horizon is crossed twice and each time
$at'\rightarrow at'-i\pi/2$ because of
(\ref{TR1}).
The remaining term of (\ref{TU}) gives $k_{1}\Delta x'=k_{1}x'-(-k_{1})(-x')=0$.
The final result is therefore
\begin{equation}\label{TU1}
T=\frac{a}{2\pi}\,,
\end{equation}
which is independent of $k^{1}$ and coincides with the usual Unruh temperature \cite{UN,CRIS}.
This result, with the replacement $a\rightarrow a/\sqrt{1-a^2 /\mathcal{A}^2}$,
where $\mathcal{A}=2m$ is the maximal acceleration,
also confirms a recent calculation \cite{BEFE} regarding particles whose acceleration has an upper limit.
Equation (\ref{TU1}) comes in fact from the term $k_{0}\Delta t'$ that does not contain
derivatives of $\gamma_{\mu\nu}$. The difference from \cite{BEFE}, as well as from \cite{DEG}, is however represented
by the form  of (\ref{TA}) of the decay rate \cite{AKM} which carries a factor $2$ in the exponential as required by our invariant approach.

Despite its limitations, the external field approximation
already reproduces (\ref{TU1}) at $O(\gamma_{\mu\nu})$. Additional terms of (\ref{TAG}) are expected to contain corrections to (\ref{TU1}).
We note, however, that for a closed space-time path the last integral in (\ref{TA}) and (\ref{TU}) becomes
$\int_{\Sigma} d\sigma^{\mu\lambda}R_{\mu\lambda\alpha\beta}J^{\alpha\beta}$, where $\Sigma$ is the surface bounded by the path \cite{PAP2},
and has an imaginary part if $R_{\mu\nu\alpha\beta}$ has singularities. This eventuality may call
for a complete quantum theory of gravity \cite{MISN}.

\section{$K_{\lambda}$ in interaction}

{\it i. Poynting vector.} The question we ask in this section is whether the vector $K_{\lambda}$ is redundant, or plays a role in radiation problems.
Using $\tilde{F}_{\mu\nu}$, we can construct, for instance, a "Poynting" vector.
Assuming, for simplicity, that $j_{\mu}=0$ in (\ref{ME2}), using known vector
identities, integrating over a finite volume and reverting to normal units,  we obtain from (\ref{ME1}) and (\ref{ME2}) the
conservation equation
\begin{equation}\label{CON}
\frac{1}{c}\frac{\partial}{\partial t}\int\left(\tilde{E}^{2}+\tilde{H}^{2}\right)dV= -2\oint\vec{\tilde{S}}\cdot d\vec{\Sigma}\,,
\end{equation}
where $\Sigma$ is the surface bounding $V$ and $\vec{\tilde{S}}=\vec{\tilde{E}}\times \vec{\tilde{H}}$ is the gravitational
Poynting vector. Both sides of (\ref{CON}) acquire, in fact, the dimensions of an energy flux after multiplication by $G/c^{3}$.
We can now calculate the flux of $\vec{\bar{S}}$ at the particle assuming that the momentum of the free particle is $k\equiv k^{3}$ and that the
source in $V$ emits a plane gravitational wave in the $x$-direction. In this case the wave is
determined by the components $\gamma_{22}=-\gamma_{33}$ and $\gamma_{23}$, and we find $\tilde{E}_{1}=0$, $\tilde{E}_{2}=2R_{0203}J^{03}+2R_{0231}J^{31}$,
$\tilde{E}_{3}=2R_{0303}J^{03}+2R_{0331}J^{31}$, $\tilde{H}_{1}=0$, $\tilde{H}_{2}=-4R_{3103}J^{03}-4R_{3113}J^{13}$, $\tilde{H}_{3}=4R_{2103}J^{03}+4R_{2113}J^{13}$.
It also follows that $R_{0203}=R_{0231}=R_{2103}-R_{2113}=-\ddot{\gamma}_{23}/2$ and $R_{0303}=R_{0331}=R_{3103}=R_{3113}=\ddot{\gamma}_{22}/2$.
The action of $\tilde{S}$ on the quantum particle is directed along the axis of propagation of the wave and results in a combination of oscillations and rotations
about the point $x^{\alpha}$ with angular momentum given by $2J^{03}= (x^{0}-z^{0})k-k^{0}(x^{3}-z^{3})$, $2J^{13}=(x^{1}-z^{1})k$
and $2J^{23}=(x^{2}-z^{2})k$. A similar motion also occurs in the case of Zitterbewegung \cite{PAP11}.
Reverting to normal units, the energy flux associated with this process is
\begin{equation}\label{fi}
\Phi = (\omega^{4}G/c^{3})\left\{ (\gamma^{23})^{2}[(J^{03})^{2}+J^{31}J^{03}]+(\gamma^{22})
^{2}[(J^{03})^{2}-J^{31}J^{03}-(J^{31})^{2}] \right\}
\end{equation}
and increases rapidly with the wave frequency $\omega$ and the particle's angular momentum.

{\it ii. Electromagnetic radiation.} Let us assume that a spinless particle has a charge $q$. Acceleration, whatever its cause, makes the particle radiate electromagnetic waves. The
four-momentum radiated away by the particle, while passing through the driving gravitational field $\tilde{F}_{\mu\nu}$, is given by the formula
\begin{equation}\label{MOM}
\Delta p^{\alpha}=-\frac{2q^2}{3c}\int \frac{du_{\beta}}{ds}\frac{du^{\beta}}{ds}dx^{\alpha}=
-\frac{2q^2}{3c}\int \left(\tilde{F}_{\mu\nu}u^{\nu}\right)\left(\tilde{F}^{\mu\delta}u_{\delta}\right)dx^{\alpha}\,,
\end{equation}
that can be easily expressed in terms of the external fields (\ref{EM}) on account of the equation of motion
of the charge in the accelerating field \cite{PAP10}.
At this level of approximation the particle can distinguish uniform
acceleration which gives $\Delta p^\alpha\sim \int g^2 dx^{\alpha}$, where $g$ is a constant, from a non-local
gravitational field and radiates accordingly. This is explained by the presence of $R_{\mu\nu\alpha\beta}$  in (\ref{EM}) and is a direct consequence of our use of
the equation of geodesic deviation in (\ref{MOM}).

When the accelerating field is the wave discussed above, the incoming gravitational wave and the
emitted electromagnetic wave have the same frequency $\omega$ and the
efficiency of the gravity induced production of photons increases as $\omega^{4}k^{2}$.

{\it iii. Flux quantization.} Flux quantization is the typical manifestation
of processes in which the wave function is non-integrable. Of interest is here the presence of the free particle momentum $k^{\alpha}$
in $K_{\lambda}$.

Let us consider for simplicity the case of a
rotating superfluid. Then $\gamma_{01}=-\Omega z_{2}/c$, $\gamma_{02}=\Omega z_{1}/c$ and the remaining metric components vanish.
The angular velocity $\Omega$ is assumed to be constant in time and $k_{3}=0$. Without loss of generality, we can also choose the reference point $x^{\mu}=0$.
We find $K_{0}=K_{3}=0$ and
\begin{equation}\label{KR}
K_{1}=-\frac{1}{2}\left[\gamma_{01,2}z^{2}-\gamma_{01}\right]k^{0}-\frac{1}{2}\left[-\gamma_{01,2}z^{0}\right]k^{2}
\end{equation}
\[K_{2}=-\frac{1}{2}\left[\gamma_{02,1}z^{1}-\gamma_{02}\right]k^{0}-\frac{1}{2}\left[-\gamma_{02,1}z^{0}\right]k^{1}\,.\]
Integrating over a loop of superfluid, the condition that the superfluid wave function be single-valued
gives the quantization condition
\begin{equation}\label{FQ}
\oint dz^{\lambda}K_{\lambda}=-\frac{\Omega z^{0}}{2c}\oint\left(k^{2}dz^{1}-k^{1}dz^{2}\right)=\frac{\pi \Omega z^{0}}{c}k\varrho=2\pi n\,,
\end{equation}
where $n$ is an integer, $k=\sqrt{k_{1}^{2}+k_{2}^{2}}$ and $\varrho =\sqrt{z_{1}^{2}+z_{2}^{2}}$.
The time integrating factor $z^{0}$, extended to $N$ loops, becomes $z^{0}=2\pi\varrho \varepsilon N/pc$, where $\varepsilon^{2}=(pc)^{2}+(mc^{2)^{2}}$ and $p=\hbar k$.
The superfluid quantum of circulation satisfies the condition
\begin{equation}\label{CR}
\Omega(\pi \varrho^{2})\varepsilon N/c^{2}= n\hbar\,.
\end{equation}
If the superfluid is charged, then the wave function is single-valued if the total phase satisfies the relation
\begin{equation}\label{SQ}
\oint dz^{\lambda}K_{\lambda}+\frac{q}{c}\oint dz^{\lambda}A_{\lambda}=2\pi n\hbar\,,
\end{equation}
which, for $n=0$ and zero external magnetic field, leads to $\int \vec{H}\cdot d\vec{\Sigma}=-2\pi^{2}\Omega\varrho^{2}\varepsilon N/qc$.
In this case, therefore, rotation generates a magnetic flux through $\Sigma$ and, obviously, a current in the $N$ superconducting loops.
No fundamental difference is noticed in this case from DeWitt's original treatment of the problem \cite{DEW1,PAP66,BAKK,CAS}.
\section{ Conclusions}

The two-point potential $K_{\mu}(z,x)$ plays a prominent role in the solution of
covariant wave equations through the phase $\Phi_{G}$ .
It satisfies Maxwell-type equations identically, depends on the metric tensor and is complementary to it.
The potential $K_{\lambda}$ suggests the introduction of the notion of quasiparticle
because gravity affects in general the dispersion relations of the particles with which it
interacts, as shown by (\ref{eff}), and because it carries with itself information
about matter through the particle momentum $k_{\alpha}$.

Some particular aspects of the behaviour of $K_{\lambda}$ have been examined. We have found that
when $j_{\mu}=0$, scale invariance assures that a gas of gravitons satisfies Planck's
radiation law, but that this is no longer so, in principle, for non-pure
gravitational fields.

$K_{\lambda}$ also determines the equations of motion of a particle through (\ref{tool1}),
(\ref{tool2}), (\ref{acc}) and (\ref{IT}). We have found that the motion follows
a geodesic only if the quantum particle chooses, among all available paths, that
for which $Dk_{\alpha}/Ds =0$. Along this particular path the principle of equivalence is obviously satisfied.
We have then shown that the particle motion is contained in the solution (\ref{PHA}) of
the covariant KG equation.

We have also studied quantum mechanical tunneling through a horizon and derived a covariant
and canonical invariant expression for the transition amplitude. Though the external field
approximation looks ill-suited to deal with regions of space-time close to a gravitational horizon,
the approximation reproduces the Unruh temperature exactly in the case of the Rindler metric.
No corrections and no effects due
to $k_{\mu}$ have been found to the standard result to $O(\gamma_{\mu\nu})$.
Higher order approximations can be calculated by applying (\ref{TU}).

Because $\tilde{F}_{\mu\nu}$ satisfies Maxwell-like equations, it is also possible to define a Poynting vector
and a flux of
energy and angular momentum at the particle so that the particle's motion
can be understood as a sequence of oscillations and rotations similar to what found in the case
of Zitterbewegung \cite{PAP11}.

Use of $ K_{\lambda}$ in problems where gravity accelerates a charged particle and
electromagnetic radiation is produced offers
a rather immediate relationship between the loss of energy-momentum by the quantum particle and
the driving gravitational field. These processes could give sizeable contributions
for extremely high values of $\omega$. Astrophysical processes like photoproduction \cite{PVAL}
and synchrotron radiation \cite{MIS} have been discussed in the literature
and are worthy of re-consideration in view of the present results.
An advantage on the high frequency detection side, for which detection schemes are
in general difficult to conceive, is
represented by the efficiency of the graviton-photon conversion rate
and by the high coupling afforded by a radio receiver over, for instance, a mechanical one.
This would enable, in principle, a spectroscopic analysis of the signal.

In the last problem considered, we have calculated the flux of $K_{\lambda}$ in the typical
quantum case of a non-integrable wave function. Here too, it is possible to isolate
quantities of physical interest, like magnetic flux, or circulation, despite the
non-intuitive character of $\oint dz^{\lambda}K_{\lambda}$. Unlike \cite{DEW}, our procedure
and results are fully relativistic. They can be applied directly to boson condensates in boson stars \cite{PIER}.




\end{document}